\documentclass[aps, a4paper,superscriptaddress, nofootinbib,10pt]{revtex4}
\usepackage{amsmath,amssymb, bm, natbib}
\usepackage{dsfont}
\usepackage{epsfig}
\usepackage{graphicx}
\usepackage{slashed}
\usepackage{color}
\usepackage{hyperref}
\usepackage{pstricks}
\usepackage{accents}
\usepackage{mathrsfs}
\usepackage[center,footnotesize,hang]{subfigure}
\usepackage{bigstrut}
 \usepackage[normalem]{ulem}
 
\begin{document}

\title{ 750 GeV Diphoton excess from Gauged $B - L$ Symmetry }

\author{Tanmoy Modak}
\email{tanmoyy@imsc.res.in}
\affiliation{The Institute of Mathematical Sciences, IV Cross Road, CIT Campus, Taramani, Chennai 600113, India} 
\author{Soumya Sadhukhan}
\email{soumyasad@imsc.res.in}
\affiliation{The Institute of Mathematical Sciences, IV Cross Road, CIT Campus, Taramani, Chennai 600113, India} 
\author{Rahul Srivastava}
\email{rahuls@imsc.res.in}
\affiliation{The Institute of Mathematical Sciences, IV Cross Road, CIT Campus, Taramani, Chennai 600113, India}


\begin{abstract}
   \vspace{1cm}
   
  \begin{center}
\textbf{ Abstract} 
\end{center}

We show that the recently observed 750 GeV diphoton excess at LHC can be due to the decay of a $SU(2)_L$ singlet scalar particle having 3 units of charge under gauged $B - L$ symmetry.  Such a particle arises as an essential ingredient of recently studied gauged $B -L$ extension of the Standard Model with unconventional charge assignment for right handed neutrinos. Apart from being one of the simplest extensions of the Standard Model, the model also contains a dark matter candidate and Dirac neutrinos with naturally small masses.

\end{abstract}

\maketitle


\section{INTRODUCTION}
\label{sec1}

The ATLAS and CMS collaborations at the LHC have recently reported an excess of events in the
invariant mass distribution containing two photons at $\sqrt{s}=13$
TeV~\cite{talkatlascms,atlasdiphoton,cmsdiphoton}. 
The ATLAS Collaboration~\cite{atlasdiphoton}, with 3.2 fb$^{-1}$ data, has reported
an excess of $3.9\sigma$  at diphoton invariant mass around 750 GeV. The significance
reduces to $2.3\sigma$
once the Look Elsewhere Effect is included. This corresponds to an excess in signal 
$\sigma(pp\to \gamma \gamma)$ of about ($10\pm 3$~fb) with a best fit width $\sim 45$ GeV.
The value of the experimental acceptance in ATLAS is about $0.4$. \\

The CMS collaboration has also found an excess in diphoton events with local
significance $2.6\sigma$~\cite{cmsdiphoton}
 at $\sqrt{s}=13$ TeV with 2.6 fb$^{-1}$ data at mass around $750$ GeV. This significance reduces to $2.0\sigma$ if large width 
($\sim 45$ GeV) is assumed. This  translates to an excess in signal cross section
$\sigma(pp\to \gamma \gamma)$ of
about ($6\pm 3$~fb).  \\

 These excess events do not have any significant missing energy, leptons or jets
associated with them. 
No excess of events have been found in $ZZ$, dilepton, dijet channels in the same
invariant mass region for new data.
Although this excess could well be a statistical fluctuation, it has drawn
significant attention as it can also arise from decay of new particle with mass around $750$ GeV~\cite{Models:2015Hgaga}.
\\

If the observed diphoton excess indeed corresponds to decay of a hitherto unknown particle then this will be the first confirmation of new physics beyond the Standard Model (SM). 
If the observed excess is due to a resonance it has to be a boson and it
cannot be a spin-$1$ particle~\cite{Landau:1948kw,Yang:1950rg}. This leaves the possibility of it being either a spin-$0$ or spin-$2$ particle.
If it is indeed a new particle then one must wonder what kind of new physics incorporates it. This issue can only be settled by looking at various possible 
new physics scenarios that can potentially lead to such a particle. It is generally expected that this new physics will also be related to other open problems in high energy physics which 
do not have a satisfactory explanation within SM. Chief among them is the problem of neutrino masses (and their relative smallness) and the nature of dark matter. It will be quite satisfying if the observed new particle has a natural connection with the models addressing at least one of these issues.  In this work we assume this resonance to be
a spin-$0$ particle and look at one such promising model based on gauge $B - L$ symmetry. This model was recently proposed to explain the smallness of neutrino mass if neutrinos
are Dirac particles and also has an interesting dark matter candidate
\cite{Ma:2014qra, Ma:2015mjd, Ma:2015raa}.  \\

The gauged $B- L$ symmetry is one of the simplest and most well studied extension of SM \cite{Marshak:1979fm, Davidson:1978pm}. In SM, Baryon number $B$ and Lepton number $L$ are accidentally conserved classical symmetries. However, both $B$ and $L$ currents are anomalous and only the combination $B - L$ is anomaly free. In the conventional gauged $B -L$ model,  the $B - L$ symmetry is promoted to an anomaly free gauge symmetry by addition of three right handed neutrinos $\nu^i_R$ each transforming as $-1$ under the $U(1)_{B - L}$ \cite{Marshak:1979fm, Davidson:1978pm}. It was shown that if this $B - L$ symmetry is spontaneously broken by a $SU(2)_L$ singlet scalar $\chi_2$ having two units of $B - L$ charge, then the right handed neutrinos can acquire a Majorana mass term $M_R$ proportional to the vacuum expectation value (vev) $u_2$ of the singlet scalar. Moreover if the $B - L$ breaking scale is far greater than the electroweak scale then the right handed neutrinos acquire a large mass
leading to a natural implementation of Type I seesaw mechanism. However, in this scenario the $B - L$ breaking scale is expected to be very high (same as seesaw scale) and it 
is very difficult to test this model at LHC. \\

Recently, another simple choice of $B - L$ charges for right handed neutrinos which leads to anomaly free $U(1)_{B-L}$ gauge symmetry has been proposed.  Unlike the previous case, here the three right handed neutrinos transform as $\nu^i_{R} = (+5, -4, -4)$ under $B-L$ symmetry \cite{Ma:2014qra, Montero:2007cd}. It was shown that such a charge assignment can lead to Dirac neutrinos with naturally small masses if the $B - L$ symmetry is spontaneously broken by a $SU(2)_L$ singlet scalar $\chi_3$ transforming as $\sim 3$ under $U(1)_{B-L}$ symmetry. This new $B - L$ model can also have a candidate for long lived scalar dark matter if two new singlet scalars transforming as $\chi_2 \sim 2$ and $\chi_6 \sim -6$ under $U(1)_{B-L}$ symmetry are added to it \cite{Ma:2015mjd}. Unlike the conventional $B - L$ model, here the $B - L$ breaking scale need not be high and can be well within the reach of LHC. This opens up the possibility of testing various features of this model in the present run of LHC. Moreover, the 
dark matter in this model has a significant interaction with the nuclei and can be detected in present or near future dark-matter direct-search experiments. We refer the interested readers to \cite{Ma:2014qra, Ma:2015mjd, Ma:2015raa} for further details. Thus, apart from providing a explanation for the nature and small masses for neutrinos as well as a candidate for dark matter, various aspects of this new $B - L$ model are quite testable both in colliders and dark matter direct detection experiments. Furthermore, as we will discuss in the subsequent sections, owing to presence of singlet scalars, this model is ideally suited to explain the recently observed $750$ GeV diphoton excess and the aim of this paper is to look at this possibility in details.  \\

 The plan of the paper is as follows. In Section \ref{sec2} we look at the details of the gauged $B - L$ symmetry model which is a slightly extended version of the previously discussed 
 model. We show that the modified model is also free of anomalies.  In Section \ref{sec3} we look at the details of the scalar and Yukawa sectors and identify a viable candidate which can  explain the observed $750$ GeV diphoton excess.  In Section \ref{sec4} we discuss the production and decay of the $750$ GeV particle and compare our computation with the experimental results. We finally conclude in \ref{sec5}.


\section{The Gauge $B - L$ Symmetry Model}
\label{sec2}


 The anomaly free gauged $B - L$ model with unconventional $B - L$ charges for right handed neutrinos was originally constructed to obtain Majorana neutrinos \cite{Montero:2007cd} or Dirac neutrinos \cite{Ma:2014qra,Ma:2015raa} with naturally small masses. The model for Dirac neutrinos was further extended in \cite{Ma:2015mjd} to accommodate a long lived dark  matter particle. The $SU(2)_L$ singlet scalars of the model are required to break the gauge $B - L$ symmetry and we show in this work, that a linear combination of these scalar can be a viable candidate for $750$ GeV resonance, whose decay can provide a possible explanation for the recently observed diphoton excess \cite{atlasdiphoton,cmsdiphoton}.  In this work we study a slightly extended version of the model discussed in \cite{Ma:2015mjd}, where we have also included two $SU(2)_L$ singlet vector ``quarks'' $X_{L,R}, Y_{L,R}$. Although they are $SU(2)_L$ singlets, these exotic quarks do carry $SU(3)_c$ colour charge as well as $U(1)_Y$, $U(1)_{B-L}$ charges.   The $SU(3)_c \times SU(2)_L \times U(1)_
Y$ and $U(1)_{B - L}$ charge assignment for the fermions in the model are as follows:
 
  \begin{table}[h]
\begin{center}
\begin{tabular}{c c c || c c c}
  \hline \hline
  Fields                      \hspace{0.2cm}                    & $SU(3)_c \times SU(2)_L \times U(1)_Y$                \hspace{0.2cm}                                    &  $U(1)_{B - L}$              \hspace{0.2cm}              & 
  Fields                      \hspace{0.2cm}                    & $SU(3)_c \times SU(2)_L \times U(1)_Y$                \hspace{0.2cm}                                    &  $U(1)_{B - L}$                                          \\
  \hline \hline
  $Q^i_L$                     \hspace{0.2cm}                    & $(3, 2, \frac{1}{3})$                                 \hspace{0.2cm}                                    &  $\frac{1}{3}$               \hspace{0.2cm}               &   
  $L^i_L$                     \hspace{0.2cm}                    & $(1, 2, -1)$                                          \hspace{0.2cm}                                    &  $-1$                                                     \\
  $u^i_R$                     \hspace{0.2cm}                    & $(3, 1, \frac{4}{3})$                                 \hspace{0.2cm}                                    &  $\frac{1}{3}$               \hspace{0.2cm}                &   
  $l^i_R$                     \hspace{0.2cm}                    & $(1, 1, -2)$                                          \hspace{0.2cm}                                    &  $-1$                                                      \\
  $d^i_R$                     \hspace{0.2cm}                    & $(3, 1, -\frac{2}{3})$                                \hspace{0.2cm}                                    &  $\frac{1}{3}$               \hspace{0.2cm}                 & 
  $\nu^1_{R} $                \hspace{0.2cm}                    & $(1, 1, 0)$                                           \hspace{0.2cm}                                    &  $5$                                                       \\
  $\nu^2_{R} $                \hspace{0.2cm}                    & $(1, 1, 0)$                                           \hspace{0.2cm}                                    &  $-4$                        \hspace{0.2cm}                 & 
  $\nu^3_{R} $                \hspace{0.2cm}                    & $(1, 1, 0)$                                           \hspace{0.2cm}                                    &  $-4$                                                      \\
  $N^i_L$                     \hspace{0.2cm}                    & $(1, 1, 0)$                                           \hspace{0.2cm}                                    &  $-1$                        \hspace{0.2cm}                 &
  $N^i_R$                     \hspace{0.2cm}                    & $(1, 1, 0)$                                           \hspace{0.2cm}                                    &  $-1$                        \hspace{0.2cm}                 \\
  $X_L$                       \hspace{0.2cm}                    & $(3, 1, \frac{4}{3})$                                 \hspace{0.2cm}                                    &  $3$                         \hspace{0.2cm}                 &
  $X_R$                       \hspace{0.2cm}                    & $(3, 1, \frac{4}{3})$                                 \hspace{0.2cm}                                    &  $0$                         \hspace{0.2cm}                 \\
  $Y_L$                       \hspace{0.2cm}                    & $(3, 1, -\frac{4}{3})$                                \hspace{0.2cm}                                    &  $-3$                        \hspace{0.2cm}                 &
  $Y_R$                       \hspace{0.2cm}                    & $(3, 1, -\frac{4}{3})$                                \hspace{0.2cm}                                    &  $0$                         \hspace{0.2cm}                 \\
  \hline
  \end{tabular}
\end{center}
\caption{The $SU(3)_c \times SU(2)_L \times U(1)_Y$  and $U(1)_{B-L}$ charge assignment for the fermions. Here $i = 1, 2, 3$ represents the three generations.}
  \label{tab1}
\end{table} 

 In Table \ref{tab1} apart from the SM particles we have also included three right handed neutrinos $\nu^i_R$, three $SU(2)_L$ singlet heavy fermions $N^i_{L,R}$ (as in the previous model \cite{Ma:2015mjd}) and two pair of exotic ``quarks'' $X_{L,R}$, $Y_{L,R}$ which carry color and electromagnetic charges but are singlet under $SU(2)_L$.  \\
 
 The charge assignment for the scalars in this model (which are same as in \cite{Ma:2015mjd}) are as follows:
   
   \begin{table}[h]
\begin{center}
\begin{tabular}{c c c || c c c}
  \hline \hline
  Fields                      \hspace{0.1cm}                    & $SU(3)_c \times SU(2)_L \times U(1)_Y$                \hspace{0.1cm}                                    &  $U(1)_{B - L}$              \hspace{0.1cm}              & 
  Fields                      \hspace{0.1cm}                    & $SU(3)_c \times SU(2)_L \times U(1)_Y$                \hspace{0.1cm}                                    &  $U(1)_{B - L}$                                          \\
  \hline \hline
  $\Phi = (\phi^+, \phi^0)^T$ \hspace{0.1cm}                    & $(1, 2, 1)$                                           \hspace{0.1cm}                                    &  $0$                         \hspace{0.1cm}               &   
  $\chi_2$                     \hspace{0.1cm}                   & $(1,1, 0)$                                            \hspace{0.1cm}                                    &  $2$                                                     \\
  $\chi_3$                     \hspace{0.1cm}                   & $(1, 1, 0)$                                           \hspace{0.1cm}                                    &  $3$                          \hspace{0.1cm}                &   
  $\chi_6$                     \hspace{0.1cm}                   & $(1, 1, 0)$                                           \hspace{0.1cm}                                    &  $-6$                                                     \\
   \hline                                                                                                                                                                                                                             
     \end{tabular}
\end{center}
\caption{The $SU(3)_c \times SU(2)_L \times U(1)_Y$  and $U(1)_{B-L}$ charge assignment for the scalars.}
  \label{tab2}
\end{table}  

 In Table \ref{tab2}, $\Phi = (\phi^+, \phi^0)^T$ is the usual $SU(2)_L$ doublet scalar and $\chi_i$ are $SU(2)_L$ singlet scalars. The new fermions introduced in the model can potentially lead to anomalies. Thus, it is important to ensure that the model is anomaly free. The new particles can induce following triangular anomalies:

 \begin{eqnarray}
[SU(3)_c]^2 \, U(1)_{B-L}                \quad   & \rightarrow &  \quad     \sum_q (B-L)_{q_L}  - \sum_q (B-L)_{q_R}       \\
\left[SU(2)_L\right]^2   \, U(1)_{B-L}   \quad   & \rightarrow &  \quad      \sum_l (B-L)_{l_L}  + 3 \, \sum_q (B-L)_{q_L}     \\
\left[U(1)_Y\right]^2   \, U(1)_{B-L}    \quad   & \rightarrow &  \quad      \sum_{l,q} \left[ Y^2_{l_L} \, (B-L)_{l_L}  + 3 \, Y^2_{q_L} \,  (B-L)_{q_L} \right]  
\, - \,  \sum_{l,q} \left[ Y^2_{l_R} \, (B-L)_{l_R}  + 3 \, Y^2_{q_R} \,  (B-L)_{q_R} \right]  \\
U(1)_Y \, \left[U(1)_{B-L}\right]^2     \quad   & \rightarrow &  \quad      \sum_{l,q} \left[ Y_{l_L} \, (B-L)^2_{l_L}  + 3 \, Y_{q_L} \,  (B-L)^2_{q_L} \right]  
\, - \,  \sum_{l,q} \left[ Y_{l_R} \, (B-L)^2_{l_R}  + 3 \, Y_{q_R} \,  (B-L)^2_{q_R} \right]  \\
\left[U(1)_{B-L}\right]^3     \quad   & \rightarrow &  \quad      \sum_{l,q} \left[ (B-L)^3_{l_L}  + 3 \, (B-L)^3_{q_L} \right]  
\, - \,  \sum_{l,q} \left[ (B-L)^3_{l_R}  + 3 \, (B-L)^3_{q_R} \right]  \\
\left[\rm{Gravity}\right]^2   \, \left[U(1)_{B-L}\right]     \quad   & \rightarrow &  \quad      \sum_{l,q} \left[ (B-L)_{l_L}  + 3 \, (B-L)_{q_L} \right]  
\, - \,  \sum_{l,q} \left[ (B-L)_{l_R}  + 3 \, (B-L)_{q_R} \right]
  \label{anomaly}
 \end{eqnarray}

 It has already been shown in \cite{Ma:2014qra} that for the case when exotic quarks $X, Y$ are not present, the model is completely anomaly free. It can also be easily seen that 
 the addition of the $X, Y$ quarks does not spoil the anomaly cancellation and hence the model remains anomaly free.


\section{The Scalar and Yukawa Sectors}
\label{sec3}


In this section we look at the details of the scalar and Yukawa sector of our model and identify the candidate for $750$ GeV resonance. The scalar potential of our model is given by

\begin{eqnarray}
V & = & - \mu^2_0(\Phi^{\dagger}\Phi) + m_2^2(\chi_2^*\chi_2) - \mu_3^2(\chi_3^*\chi_3) - \mu_6^2(\chi_6^*\chi_6) + \dfrac{1}{2} \lambda_0 (\Phi^{\dagger}\Phi)^2  
+  \dfrac{1}{2} \lambda_2 (\chi_2^{*}\chi_2)^2  +  \dfrac{1}{2} \lambda_3 (\chi_3^{*}\chi_3)^2  +  \dfrac{1}{2} \lambda_6 (\chi_6^{*}\chi_6)^2  \nonumber \\
& + & \lambda_{02} (\chi_2^{*}\chi_2)(\Phi^{\dagger}\Phi) + \lambda_{03} (\chi_3^{*}\chi_3)(\Phi^{\dagger}\Phi) + \lambda_{06} (\chi_6^{*}\chi_6)(\Phi^{\dagger}\Phi)
 + \lambda_{23} (\chi_2^{*}\chi_2)(\chi_3^{*}\chi_3) + \lambda_{26} (\chi_2^{*}\chi_2)(\chi_6^{*}\chi_6) \nonumber \\
& + & \lambda_{36} (\chi_3^{*}\chi_3)(\chi_6^{*}\chi_6) + [\dfrac{1}{2} f_{36} (\chi_3^{2}\chi_6) + h.c.] + [ \dfrac{1}{6} \lambda^{\prime}_{26} (\chi_2^{3}\chi_6)+h.c.] ,
\label{scapot}
\end{eqnarray}

The Minimum of $V$ is given by:\\
\begin{equation}\label{eq2}
V_{0}=  -\mu^2_0 v^2 -\mu_3^2 u_3^2 - \mu_6^2 u_6^2 + \lambda_0 \dfrac{v^4}{2} + \lambda_3 \dfrac{u_3^4}{2} + \lambda_6 \dfrac{u_6^4}{2} + \lambda_{03}u_3^2v^2 
+ \lambda_{06} u_6^2 v^2 + \lambda_{36} u_3^2 u_6^2 + f_{36}\dfrac{u_3^2 u_6}{2} ,
\end{equation}

where  $\left<\phi^0\right> =  v,  \left<\chi_3\right> = u_3,  \left<\chi_6\right> = u_6,$   are the vev of the scalar fields. Moreover, just like in \cite{Ma:2015mjd}, here also the singlet scalar $\chi_2$ does not acquire any vev i.e. $\langle \chi_2 \rangle = 0$.  The minimum of $V$ is determined by

\begin{eqnarray}
\mu_0^2 &=& \lambda_0 v^2 + \lambda_{03} u_3^2 + \lambda_{06} u_6^2, \\ 
\mu_3^2 &=& \lambda_3 u_3^2 + \lambda_{03} v^2 + \lambda_{36} u_6^2 
+ f_{36} u_6, \\ 
\mu_6^2 &=& \lambda_6 u_6^2 + \lambda_{06} v^2 + \lambda_{36} u_3^2 
+ {f_{36} u_3^2 \over 2 u_6}. 
\end{eqnarray}

Since $\langle \chi_2 \rangle = 0$, there is one dark-matter scalar boson $\chi_2$ with mass given by
\begin{equation}
m_{\chi_2}^2 = m_2^2 + \lambda_{02} v^2 + \lambda_{23} u_3^2 + \lambda_{26} 
u_6^2.
\end{equation}

There is also one physical pseudoscalar boson

\begin{equation}
A = \sqrt{2} Im(2 u_6 \chi_3 + u_3 \chi_6)/\sqrt{u_3^2 + 4 u_6^2}
\end{equation}

with mass given by

\begin{equation}
m_A^2 = - f_{36} (u_3^2 + 4 u_6^2)/2u_6.
\end{equation}

There are three physical scalar bosons spanning the basis $[h, \sqrt{2} 
Re(\chi_3), \sqrt{2} Re(\chi_6)]$, with $3 \times 3$ mass-squared matrix 
given by

\begin{equation}
M^2 = \begin{pmatrix} 2 \lambda_0 v^2 & 2 \lambda_{03} u_3 v & 2 \lambda_{06} u_6 v 
\cr  2 \lambda_{03} u_3 v & 2 \lambda_3 u_3^2 & 2 \lambda_{36} u_3 u_6 + 
f_{36} u_3 \cr 2 \lambda_{06} u_6 v & 2 \lambda_{36} u_3 u_6 + 
f_{36} u_3 & 2 \lambda_6 u_6^2 - f_{36} u_3^2/2 u_6
\end{pmatrix}
\label{mmat}.
\end{equation}


 \subsection{Simplifying Scenario}
 \label{subsec3-1}

 The mass matrix in Eq. \ref{mmat} can be diagonalized to give three CP even scalars which will be linear combinations of $\Phi, \chi_3, \chi_6$ scalars. However for sake of illustration,  we look at a special case of the generic mass matrix in Eq. \ref{mmat} which takes a simple form if we assume
 
  \begin{eqnarray}
   2 \lambda_0 v^2 & = & a^2   \quad \quad \Rightarrow \lambda_0 \, = \, \dfrac{a^2}{2 v^2} \nonumber \\
 4 \lambda_{03} u_3 v & = & a b  \qquad \Rightarrow \lambda_{03} \, = \,  \dfrac{a b}{4 u_3 v} \nonumber\\
   4\lambda_{06} u_6 v & = & a b  \qquad \Rightarrow \lambda_{06} \, = \,  \dfrac{a b}{4 u_6 v} \nonumber\\
    2 \lambda_3 u^2_3 & = & b^2   \qquad \Rightarrow   \lambda_3 \, = \,  \dfrac{b^2}{2 u^2_3}  \nonumber\\
 4 \lambda_{36} u_3 u_6  \, + \,  2 f_{36} u_3  & = & b^2 \qquad \Rightarrow f_{36} \, = \, \dfrac{1}{2\, u_3} \left( b^2 - 4 \lambda_{36} u_3 u_6 \right) \nonumber \\
   2 \lambda_6 u^2_6 - \dfrac{f_{36} u^2_3}{2 u_6} & = & b^2 \qquad \Rightarrow \lambda_6 \, = \, \dfrac{1}{2 u^2_6} \left( b^2 \, + \, \dfrac{f_{36} u^2_3}{2 u_6} \right)
   \label{assm}
  \end{eqnarray}

  where $a$ and $b$ are two independent parameters.  With these simplifying assumptions, the mass matrix of Eq. \ref{mmat} becomes

\begin{equation}\label{shmat}
\begin{pmatrix}
a^2                 \ & \      \dfrac{a b}{2}     \ & \     \dfrac{a b}{2}       \\ \\
\dfrac{a b}{2}      \ & \       b^2               \ & \      \dfrac{b^2}{2}      \\ \\
\dfrac{a b}{2}      \ & \     \dfrac{b^2}{2}      \ & \      b^2                   \\
\end{pmatrix}
\end{equation}

The eigenvalues of the mass matrix in Eq. \ref{shmat} are given by

 \begin{eqnarray}
   \Sigma_1 & = & \dfrac{1}{4}\, \left( 2 a^2 \, + \, 3 b^2 \, - \, \sqrt{4 a^4 \, - \, 4 a^2 b^2 \, + \, 9 b^4} \right) \nonumber \\
  \Sigma_2 & = & \dfrac{b^2}{2} \nonumber \\
   \Sigma_3 & = & \dfrac{1}{4}\, \left( 2 a^2 \, + \, 3 b^2 \, + \, \sqrt{4 a^4 \, - \, 4 a^2 b^2 \, + \, 9 b^4} \right)  
  \label{egv}
 \end{eqnarray}

  The masses of the scalars are then given by

 \begin{eqnarray}
m_1 \, = \,  \sqrt{2 \Sigma_1} \, , \quad  m_2 \, = \,  \sqrt{2  \Sigma_2}\, , \quad  
m_3 \, = \,  \sqrt{2   \Sigma_3} 
  \label{egma}
 \end{eqnarray}

with one scalar having the mass same as the $125$ GeV resonance and another having mass $750$ GeV. For sake of definiteness we will identify the first eigenstate with the $125$ GeV scalar (henceforth called ``Higgs'') and the second eigenstate as $750$ GeV scalar  i.e. we demand $m_1 = 125$ GeV and $m_2 = 750$ GeV. The mass of the third scalar then depends on the value of $a$ and $b$. Solving for $a$ and $b$ we find that $a = 108.5$ GeV and $b = 750$ GeV leads to desired masses for $m_1$ and $m_2$ scalars.
The mass of the third scalar $m_3$ then becomes $m_3 = 1.30$ TeV.        \\

The masses of the pseudoscalar, dark matter and $Z'$ are dependent on the value of other free parameters e.g. the value of vevs $u_3, u_6$, the $U(1)_{B-L}$ coupling $g_X$ as well as on the quartic coupling of scalars $\lambda_{ij}$. We also like to note that in the simplified mass matrix of Eq. \ref{shmat}, not all of $\lambda_{ij}$ are independent parameters owing to Eq. \ref{assm}.  The mass of dark matter $\chi_2$ is also dependent on the additional parameter $m^2_2$ and the quartic couplings $\lambda_{i2}$; $i = 0, 3, 6$. Thus for a large range of parameter space, we can also have heavy $Z'$ as required by constraints from LUX dark matter direct detection experiment \cite{Ma:2015mjd,Akerib:2013tjd}. Since the mass of the dark matter $m_{\chi_2}$ depends on additional free parameters therefore it can either be greater than or less than $750/2 = 375$ GeV. This leads to two distinct cases; if   $m_{\chi_2} \leq 375$ GeV then the 750 GeV resonance can decay into dark matter and it can lead to 
significant invisible decay width. If  $m_{\chi_2} > 375$ GeV then this decay is kinematically forbidden. In later sections we will study both these cases in details.


\subsection{Higgs Coupling to SM Gauge Bosons}
\label{subsec3-2}


 In Section \ref{subsec3-1}, we showed that in a simplified mass matrix, one of the scalar combinations can be identified with the $125$ GeV Higgs recently discovered at LHC.
 The recent data from both ATLAS and CMS experiments suggest that this scalar has couplings very similar to SM Higgs couplings with the SM gauge bosons. In this section we show that in our model also, there exists a decoupling limit where one of the scalars will have almost SM like couplings with the SM gauge bosons.

 For sake of simplicity we will work with the simplified mass matrix of Eq. \ref{shmat}. The mass matrix mixes $\phi^R_0, \chi^R_3, \chi^R_6$ states with each other. This matrix can be diagonalized by an orthogonal matrix with the diagonal matrix corresponding to the masses of the three physical scalars as shown in Eq. \ref{egv}. The physical scalars are then given by
 
 \begin{eqnarray}
    h & = &  \cos \theta \, \phi^R_0 \, - \, \sin \theta (\chi^R_6 \, + \, \chi^R_3) \, = \, 125 \, \rm{GeV}  \nonumber \\
    H_1 & = &  (\chi^R_6 \, - \, \chi^R_3)  \, = \, 750 \, \rm{GeV}  \nonumber \\
    H_2 & = & \sin \theta \, \phi^R_0 \, + \, \cos \theta (\chi^R_6 \, + \, \chi^R_3)
  \label{phys}
 \end{eqnarray}

 where $\tan 2 \theta = \dfrac{2\sqrt{2} a b}{3 b^2  - 2 a^2}$. It is clear from Eq. \ref{phys} that as $\sin \theta \rightarrow 0$, $h$ couplings to SM gauge bosons become SM like. For the case of $a = 108.5$ GeV and $b = 750$ GeV we find that $\cos \theta = 0.997$ $\sin \theta = 0.069$. This implies that in our model the couplings of the scalar $h$ of mass $125$ GeV with $W, Z$ gauge bosons are almost SM like. The small deviations from the SM like couplings are well within the experimental limits \cite{Aad:2015gba}.  
 It should be noted that in this limit the other scalars as well as the $Z'$ boson can be made heavy (assuming all couplings to be $\mathscr{O} (1)$ ) in congruence with the experimental bounds for these particles \cite{Ma:2015mjd}.


\subsection{The Yukawa Sector}
\label{subsec3-3}


   Apart from its coupling to the scalars, $\chi_3$ has following Yukawa couplings
   
   \begin{eqnarray}
     \mathcal{L}_{\chi_3} & = & f_X \, \bar{X}_L X_R \chi_3 \, + \, f_Y \, \bar{Y}_L Y_R \chi^*_3 + \, + \, f_N \, \bar{N}_L \nu^{2,3}_{R} \chi_3 \, + \, h.c. 
    \label{yuk}
   \end{eqnarray}
  
  As evident from Eq. \ref{yuk} both quarks $X, Y$ acquire mass after spontaneous breaking of $B - L$ symmetry and the masses are proportional to the vev $u_3$ of $\chi_3$. 
  The Yukawa coupling of $\chi_3$ translates into Yukawa coupling of the scalar $H_1 \equiv \chi_3 -\chi_6$.
  The Yukawa couplings of other scalars with fermions are same as in \cite{Ma:2015mjd} and we refer the interested reader to \cite{Ma:2015mjd} for further details.   
  Owing to the coupling of $\chi_3$ with quarks $X, Y$; the $750$ GeV scalar $H_1$ can be efficiently produced through gluon-gluon fusion  at LHC.  The production and decay of this scalar are discussed in details in the next section. 
   

\section{Production and Decay of $H_1$}
\label{sec4}


 In this section we look at the details of the production and decay channels for the $750$ GeV scalar at the LHC. In particular, we show that the decay of this scalar to two photons can lead to the observed diphoton excess. Moreover, as we will show, the decay of $H_1$
 to SM fermions and the Higgs are suppressed thus explaining the non-observation of any excess in other channels. \\

\begin{figure}[h!]
\includegraphics[width= 0.7\textwidth]{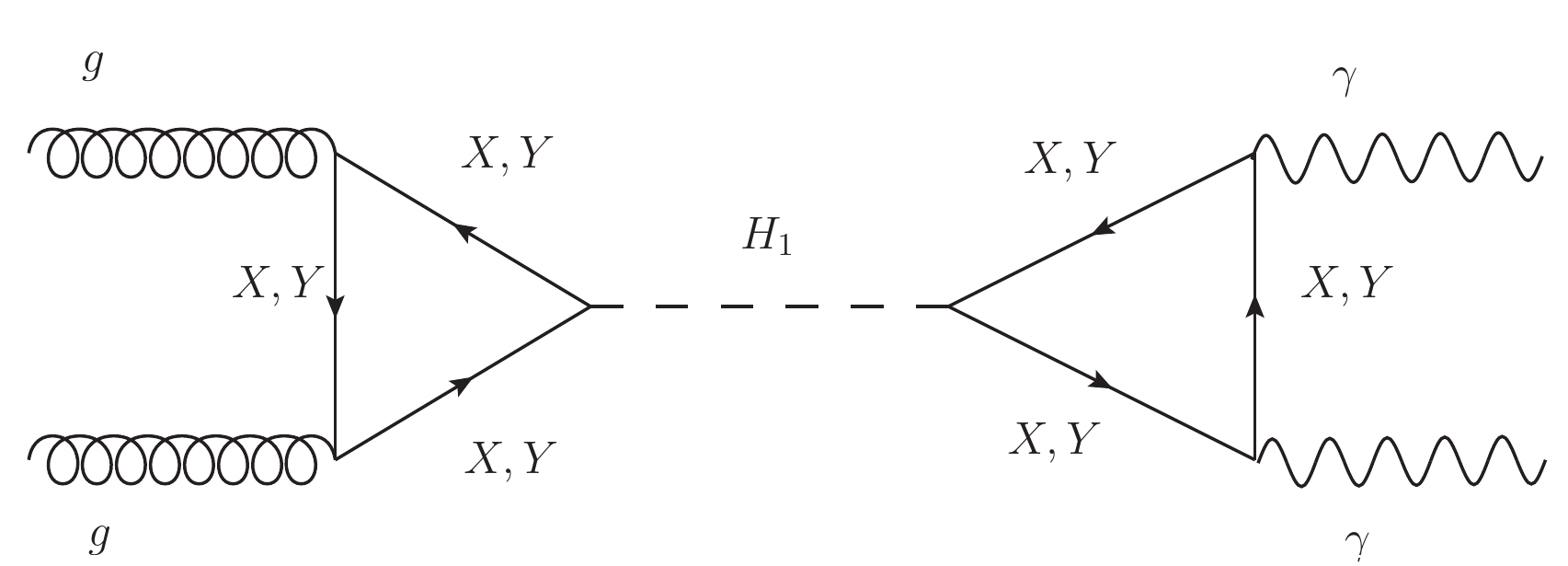}
\caption{\label{fig1} Production and decay of 750 GeV scalar to two photons.}
\end{figure}

Since, $\chi_3$ couples to quarks $X, Y$ through its Yukawa couplings Eq. \ref{yuk}, therefore, at LHC $H_1 = (\chi^R_6 \, - \, \chi^R_3)$ can be 
efficiently produced by gluon-gluon fusion through triangular loop involving $X, Y$. 
 If both $X, Y$ are heavier than $\frac{m_{H_1}}{2}$ then the tree level decay of $H_1$ to both $X, Y$ is kinematically forbidden. In such a case its decay to two photons 
through  triangular loop involving $X, Y$  as shown in fig \ref{fig1}, can be significant leading to the observed anomaly in diphoton channel. 
 
\begin{figure}[h!]
\includegraphics[width= 0.7\textwidth]{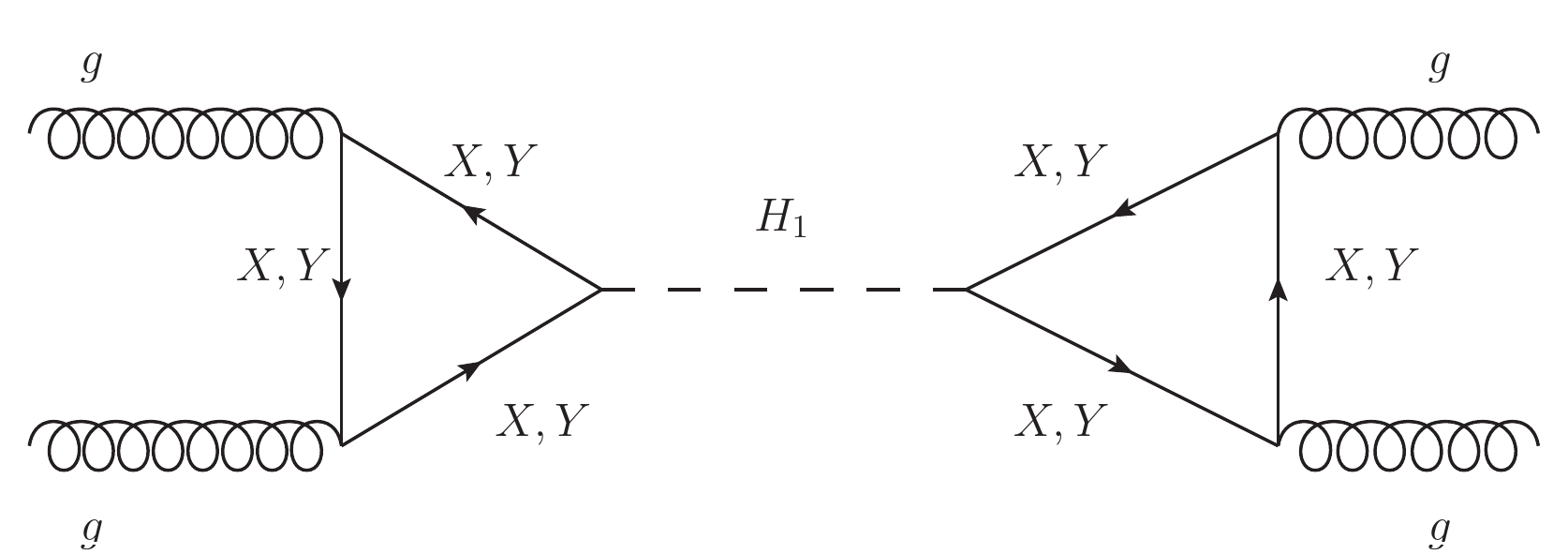}
\caption{\label{fig2} Production and decay of 750 GeV scalar to two gluons.}
\end{figure}

\begin{figure}[h!]
\includegraphics[width= 0.7\textwidth]{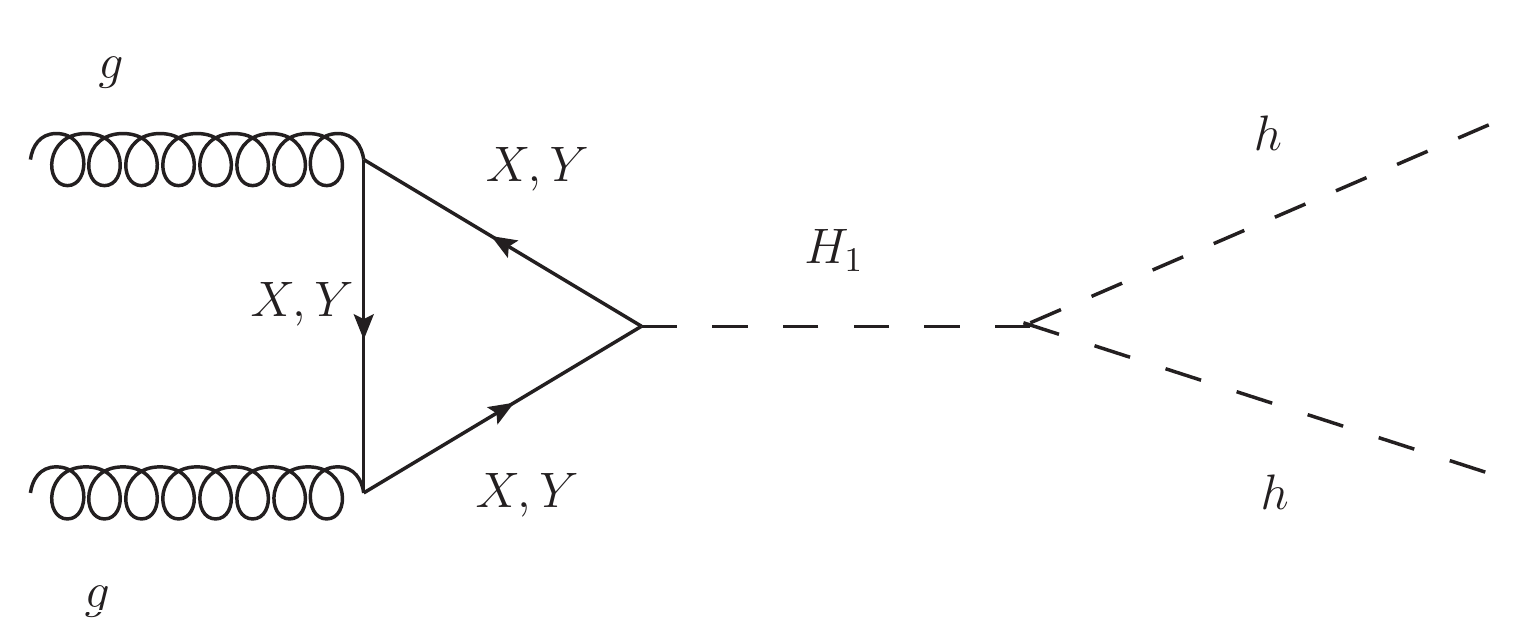}
\caption{\label{fig3} Production and decay of 750 GeV scalar to two Higgs. Owing to negligible $H_1 h h$ coupling, this decay mode is highly suppressed.  }
\end{figure}

\begin{figure}[h!]
\includegraphics[width= 0.7\textwidth]{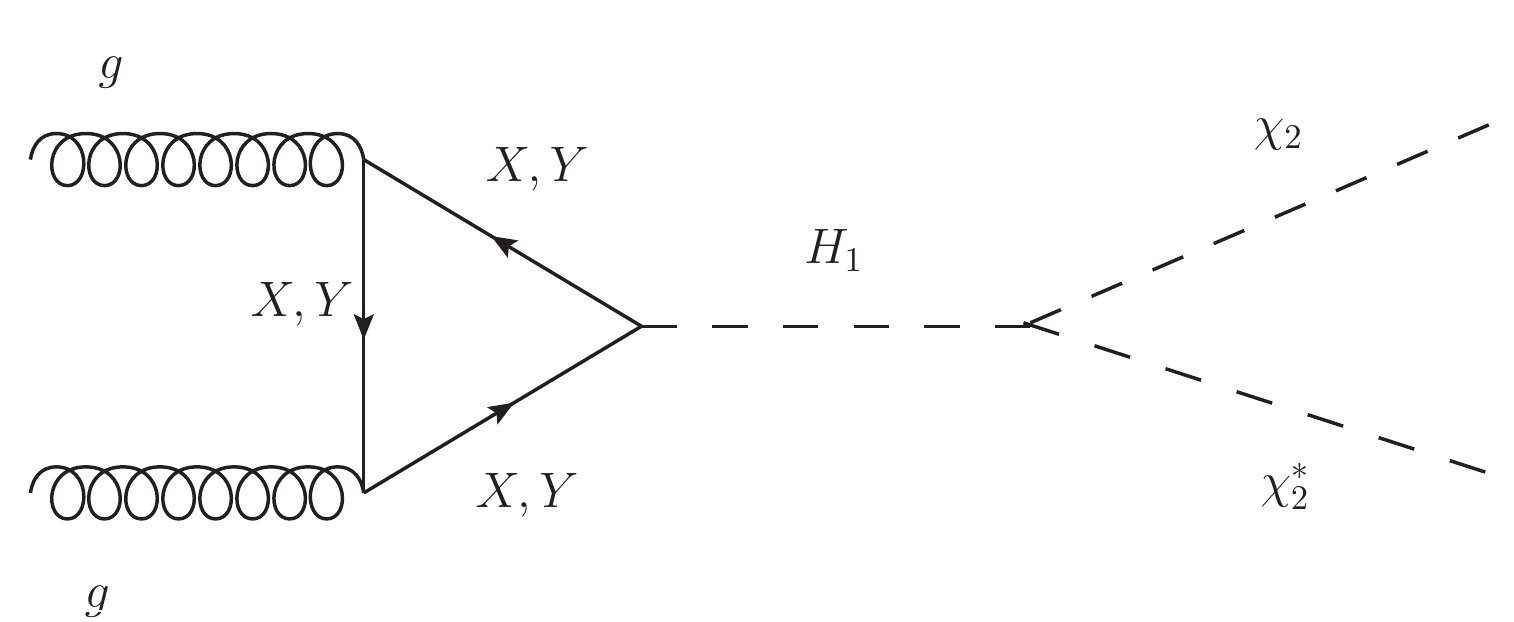}
\caption{\label{fig4} Production and decay of 750 GeV scalar to dark matter. This decay mode is only allowed if $m_{\chi_2}\leq m_{H_1}/2$.  }
\end{figure}

Apart from  its decay to two photons, $H_1$ can also decay into a pair of gluons or Higgs ($h$) as shown in fig. \ref{fig2} and fig \ref{fig3} respectively. Moreover, if the mass of dark matter $m_{\chi_2} \leq \frac{m_{H_1}}{2}$ then it can also decay into a pair of dark matter particles as shown in \ref{fig4}. This can lead to appreciable invisible decay width for $H_1$.  \\

As the SM like Higgs (h) couples to both $\chi_3$ and  $\chi_6$ with same coupling strength, this will result in cancellation of its interaction strength with $H_1 \equiv \chi_3 - \chi_6$, resulting in very small $H_1hh$ coupling.  Hence the $H_1 \to h h$ decay will be vanishingly small.  Furthermore, since $H_1$ is primarily a mixed state of $SU(2)_L$ singlets $\chi_3$ and $\chi_6$ as shown in Eq. \ref{phys} therefore it has negligible tree level coupling to SM fermions as well as $W$ and $Z$ gauge bosons. Therefore, its decay in other channels like dilepton, dijet and diboson is extremely suppressed. This observation is also in congruence with the experimental results which show lack of any statistically significant excess in these channels. \\

 Since, neither ATLAS nor CMS has seen any hint or anomalous excess in any channel for masses below $750$ GeV, we further require that all the other new particles (except dark matter) should be sufficiently massive. Thus we require that the mass of the pseudoscalar $m_A > 1$ TeV. The mass of other CP even scalar $H_2$ is also greater than 1 TeV as obtained in Section \ref{subsec3-2}. The  mass of the dark matter  $m_{\chi_2}$ is left as a free parameter and  depending on its mass, the $H_1 \to \chi_2 \chi^*_2$ decay may or may not be kinematically forbidden\footnote{Here $\chi^*_2$ denotes the anti-particle and should not be confused with an off-shell particle.}. We will consider both cases in next section. Also, we limit the mass of the $Z'$ boson $m_{Z'} > 12~$ TeV which is well above the dark matter direct detection constraints from the LUX experiment\cite{Ma:2015mjd,Akerib:2013tjd}.   \\   
                                                                                                                                                                           
Thus the only prominent decay modes of interest are $H_1 \to \gamma \gamma$, $H_1 \to g g$ and if $m_{\chi_2} \leq m_{H_1}/2$ then  $H_1 \to \chi_2 \chi_2$ also. The partial decay widths of $H_1$ in these modes are given as,
 
\begin{eqnarray}
  \Gamma (H_1 \to\gamma\gamma )&=& \frac{\alpha^2 m_{H_1}}{64 {\pi}^3} \left |2 N_c \sum_{i=X,Y} f_i Q_i^2 \sqrt{\tau_i} (1+(1-\tau_i)f(\tau_i)\right |^2,\\
 \Gamma (H_1 \to gg)&=& \frac{\alpha_s^2 m_{H_1}}{32 {\pi}^3} \left |2\sum_{i=X,Y} f_i  \sqrt{\tau_i} (1+(1-\tau_i)f(\tau_i)\right |^2, \\
  \Gamma (H_1 \to \chi_2 \chi_2) & = &    \frac{( \kappa_{\chi_2} u_3)^2}{32 \pi m_{H_1}} \left ( 1 \, - \, \frac{4 m^2_{\chi_2}}{m^2_{H_1}} \right )^{\frac{1}{2}}                                                                
  \label{phys1}
 \end{eqnarray}
 
where $\tau_i = \frac{4 m_i^2}{m_{H_1}^2}$ with $m_i, Q_i$ being corresponding fermion $(X, Y)$ masses and electromagnetic charges respectively. The $f_i~$s here denote fermion Yukawa couplings with the scalar $H_1$ whereas 
$\alpha_s, \alpha$  denote strong and electromagnetic interaction gauge coupling strengths. $N_c$ is the color factor which is 3 for the quarks.  Also without loss of generality we have normalized the dimension full coupling between $\chi_2$ and $H_1$ by the vev $u_3$ with $\kappa_{\chi_2 }$ being a dimensionless parameter. The parameter $\kappa_{\chi_2}$ is a function of the vevs $u_3, u_6$ as well as the quartic couplings between $\chi_2$ with $\chi_3$, $\chi_6$ fields as given in Eq.~\ref{scapot}. 
The $f(\tau_i)$ for our case, where  $m_{X,Y} > m_{H_1}/2$ is given as

\begin{eqnarray}
 f(\tau_i) = (\sin^{-1}[\frac{1}{\sqrt{\tau_i}}])^2.
 \label{phys2}
 \end{eqnarray}
 
   In addition to these decay modes, $H_1$ can also decay to $Z \gamma$ and $ZZ$ through triangular loops involving the $X,Y$ quarks. For our case of $m^2_Z << m^2_{H_1}$ the decay width to $Z \gamma$ and $ZZ$ are given by 
 
\begin{eqnarray} 
   \Gamma (H_1 \to Z \gamma )& = & \frac{\alpha^2 m_{H_1}}{32 {\pi}^3 s^2_W c^2_W} \left |2 N_c \sum_{i=X,Y} f_i Q_i (- Q_i s^2_W) \sqrt{\tau_i} (1+(1-\tau_i)f(\tau_i)\right |^2,\\
   \Gamma (H_1 \to Z Z )& = & \frac{\alpha^2 m_{H_1}}{64 {\pi}^3 s^4_W c^4_W} \left |2 N_c \sum_{i=X,Y} f_i (-Q_i s^2_W)^2 \sqrt{\tau_i} (1+(1-\tau_i)f(\tau_i)\right |^2  
 \label{zzcop}  
\end{eqnarray}
 
 where $s_W = \sin \theta_W$, $c_W = \cos \theta_W$ and $\theta_W$ is the electroweak angle. Since the vector quarks $X,Y$ are both $SU(2)_L$ singlets and all the three decays namely $H_1 \rightarrow \gamma \gamma, Z \gamma, Z Z$ proceed through the same triangle loop the ratio of partial decay width in these three channel are given by
 
 \begin{eqnarray} 
  \frac{ \Gamma (H_1 \to Z \gamma )}{ \Gamma (H_1 \to \gamma \gamma )} & \approx & 2 \tan^2 \theta_W, \qquad  
  \frac{ \Gamma (H_1 \to Z Z )}{ \Gamma (H_1 \to \gamma \gamma )} \, \approx \, \tan^4 \theta_W
  \label{zzratio}
 \end{eqnarray}

 As clear from Eq. \ref{zzratio}, the loop decays of $H_1$ to $Z\gamma$ and $ZZ$ are suppressed compared to the $\gamma \gamma$ decays by a factor proportional to the electroweak angle.  Thus, $H_1$ is an ideal candidate to explain the observed diphoton excess and lack of significant excess in other decay channels. 
 

\subsection{Numerical Results}
\label{subsec4-2}


 In this section we present the numerical results and the allowed parameter range for the masses of the quarks $X, Y$ and the Yukawa coupling which  can explain the excess
observed at the LHC in the diphoton channel.  For this part we have used  MadGraph5aMC@NLO\cite{Alwall:2014hca} with NN23LO1 PDF set\cite{Ball:2013hta}
to obtain the numerical estimates taking $K$ factor of 1.5 into account for NLO correction~\cite{Cline:2015msi}.\\

  For $H_1$ to be a viable candidate to explain the observed diphoton excess it not only has to explain the LHC data for $13$ TeV run but should also satisfy the nonobservance of any statistically significant excess in the previous $8$ TeV run in various channels. In our model the only significant decay channels for $H_1$ are the loop induced $gg$, $\gamma \gamma$, $Z \gamma$ and $ZZ$ decays. Moreover, if $m_{\chi_2} \leq m_{H_1}/2$ then it can also decay to two dark matter particles through $H_1 \chi_2 \chi^*_2$ tree level couplings. The $8$ TeV constraints on production $\times$ branching fraction $\sigma \times Br (h_1 \rightarrow f_i f_j)$; $f_{i,j} \equiv g,\gamma, Z, \chi_2$ on these channels are \cite{Franceschini:2015kwy, Gauge:2016jan}:
 
 \begin{eqnarray}
    \sigma \times \rm{Br}(H_1 \rightarrow \gamma \gamma) & < & 1.5 \, \rm{fb}, \qquad 
     \sigma \times \rm{Br}(H_1 \rightarrow g g) \, < \, 2500 \, \rm{fb}  , \qquad 
       \sigma \times \rm{Br}(H_1 \rightarrow \rm{invisible}) \, < \, 800 \, \rm{fb} \nonumber \\
     \sigma \times \rm{Br}(H_1 \rightarrow Z \gamma) & < & 11 \, \rm{fb}, \qquad 
     \sigma \times \rm{Br}(H_1 \rightarrow Z Z) \, < \, 12 \, \rm{fb}.   
  \label{8tevcon}
 \end{eqnarray}

As mentioned before, since in our model the dark matter mass $m_{\chi_2}$ is not fixed so there arise two distinct possibilities; either $m_{\chi_2} > m_{H_1}/2$ or $m_{\chi_2} \leq m_{H_1}/2$. For first case, $H_1$ decay to two dark matter particles is kinematically forbidden and the only prominent channels are its loop decays to $gg, \gamma \gamma$ as well as $\tan \theta_W$ suppressed loop decays to $Z \gamma$ and $Z Z$. In the second case, $H_1$ can also decay to two dark matter particles. Here we analyze both these possibilities in Subsection \ref{subsubsec4-2-1} and Subsection \ref{subsubsec4-2-2} respectively.


\subsubsection{Case-I : $m_{\chi_2} > \dfrac{m_{H_1}}{2}$}
\label{subsubsec4-2-1}


In this case  the only important decay modes for $H_1$ are $H_1 \to \gamma \gamma$ and $H_1 \to g g $ along with $H_1 \to Z \gamma $ and $H_1 \to Z Z$ both of which are $\theta_W$ suppressed. All of these decay modes are loop level, going through triangle loops involving $X, Y$ quarks. Since, $g$, $\gamma$ and $Z$ all  couple to the quarks through gauge interactions so their interaction strengths are fixed, and are proportional to $\alpha_s, \alpha$, the strong and electromagnetic coupling constants respectively. Hence, for this case of our model the production and decay rate of $H_1$ depends on only two free parameters, the masses of $X, Y$ quarks and the Yukawa coupling between $H_1$ and quarks. Moreover, since the quarks $X, Y$ acquire mass through the vev of $\chi_3$ so the Yukawa coupling can be equivalently replaced by the vev $u_3$ as a free parameter.       \\

     \begin{figure}[h!]
\includegraphics[width= 0.5\textwidth]{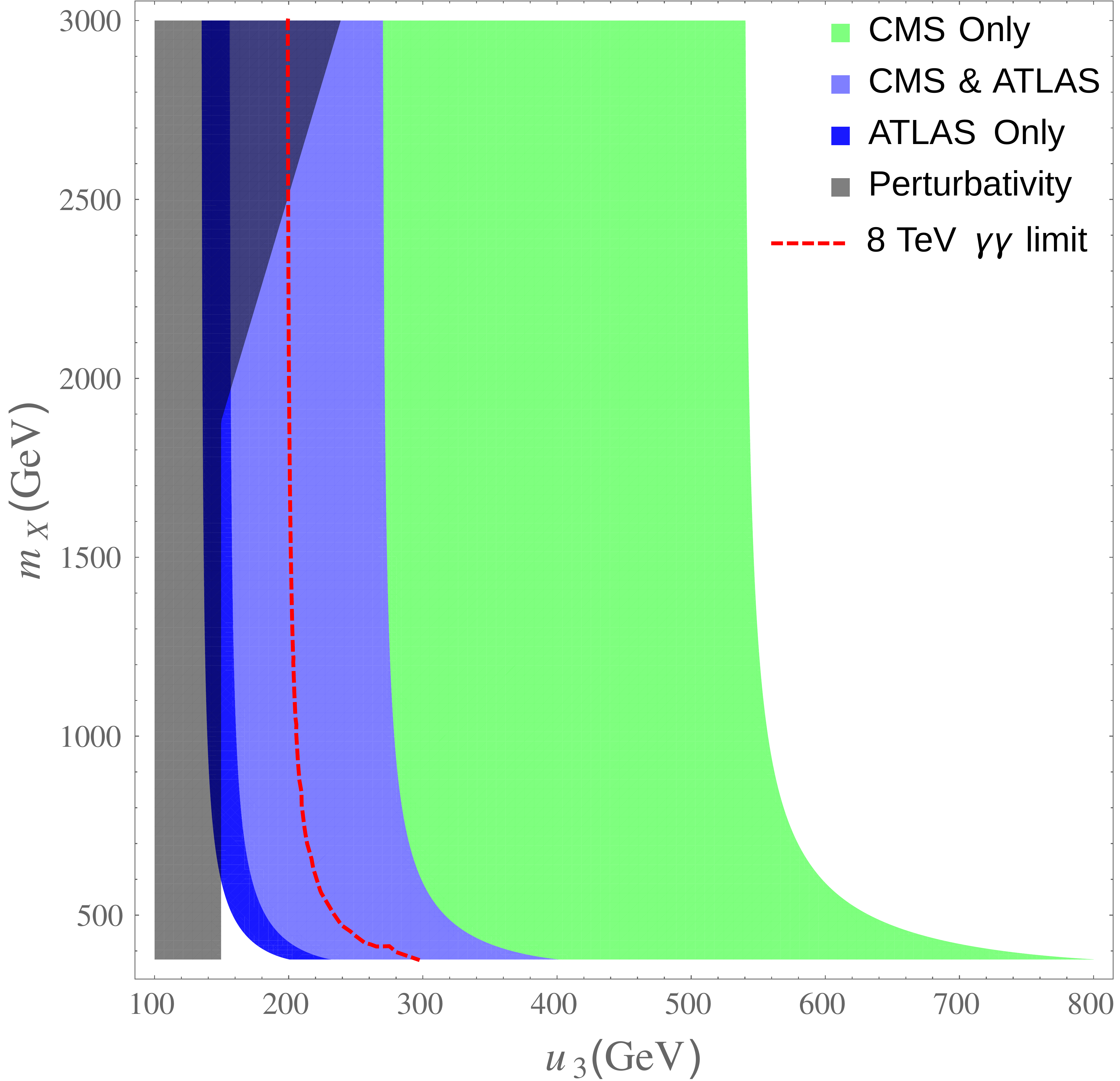}
\caption{\label{fig5} The allowed $m_X$ - $u_3$ range corresponding to CMS (green), ATLAS (deep blue) and the overlap (light blue) ranges  with
$95\%$ confidence level. Also, shown is the $95\%$ confidence level $H_1 \to \gamma \gamma$ exclusion line (red dashed) from 8 TeV run with the regions on the left of the line being incompatible with it. The black shaded region is also excluded by the perturbativity constraints.}
\end{figure}

In Fig.~\ref{fig5} we show the allowed ranges of the exotic quark masses and the
value of the vev, $\langle \chi_3 \rangle =u_3$, 
that can explain the observed 750 GeV diphoton excess for both the CMS and ATLAS
experiments within $95 \%$ confidence level. 
In obtaining the numerical results, for simplicity we assume that the masses of the exotic quarks $X,
Y$ are degenerate i.e. $m_X = m_Y$ and treat them as a single parameter $m_X$. 
In addition we require that all the couplings in our model remain perturbative. 
The region of the parameter space excluded due to non-perturbativity of the
couplings is explicitly shown in Fig.~\ref{fig5}. \\

Furthermore, in plotting Fig.~\ref{fig5} we have imposed the  8~TeV exclusion limits for the  heavy scalar of mass 750~GeV in all other channels. The strongest constraint from 8~TeV exclusion limits actually comes from non-observance of any statistically significant excess in the $\gamma \gamma$ decay channel. In Fig.~\ref{fig5} the thick red line corresponds to $\gamma \gamma$ exclusion limit of  Eq. \ref{8tevcon}. The parameter space on the left of the red line is incompatible with the 8 TeV data. \\

 As mentioned before, the scalar $H_1$ that we are considering here, does not couple to SM fermions at
tree level. Therefore the limits given in \cite{Fermi:2016jan} can  be  easily satisfied. The coupling $H_1 h h$ is also negligibly small and $\sigma(pp \to H_1 \to hh)$ is well under the
experimental limit \cite{ATLAS:2014rxa}.
Moreover, the scalar is a neutral SU(2) singlet.  Therefore, it does not have any tree level coupling to either $W$ or $Z$ bosons.  As the exotic fermions are SU(2)
singlet even the $H_1 \to WW$ decay through the triangle loop is not possible. However, it can couple to $ZZ,~ Z \gamma$ at loop level through triangle loop of the exotic fermions and it has to be taken into account. \\

Compared to the $H_1 \to \gamma \gamma$ the $ZZ,~ Z \gamma$ decays are suppressed. Furthermore, as shown in Eq. \ref{8tevcon} the exclusion limits on these decay channels are also relatively weaker. Thus the constraints from these decay channels are quite weak and do not impose any additional constraints on the allowed parameter range shown in the plot. Finally the $H_1$ decay to gluons is also well below the experimental limit and does not impose any addition constraints on the allowed parameter range. As an example the values for these decay channels for a benchmark point ($m_X = 1$ TeV and $u_3 = 205$ GeV) on the $\gamma \gamma $ exclusion line of Fig.~\ref{fig5} are given as 

 \begin{eqnarray}
   \sigma \times \rm{Br}(H_1 \rightarrow \gamma \gamma) & = & 1.5 \, \rm{fb}, \qquad 
     \sigma \times \rm{Br}(H_1 \rightarrow g g) \, = \, 490 \, \rm{fb},  \nonumber \\
     \sigma \times \rm{Br}(H_1 \rightarrow Z \gamma) & = & 0.89 \, \rm{fb}, \qquad 
     \sigma \times \rm{Br}(H_1 \rightarrow Z Z) \, = \, 0.14 \, \rm{fb}.
  \label{nodmexc}
 \end{eqnarray}

As clear from Eq.~\ref{nodmexc}, for this case of our model, apart from the $\gamma \gamma$ decay channel, the constraints from all other decay channels are easily satisfied. Even for the $\gamma \gamma$ channel, our model has enough parameter space compatible with both the observed 13~TeV excess and the 8~TeV constraints.


\subsubsection{Case-II : $m_{\chi_2} \leq \dfrac{m_{H_1}}{2}$}
\label{subsubsec4-2-2}

\begin{figure}[h!]
\includegraphics[width= 0.5\textwidth]{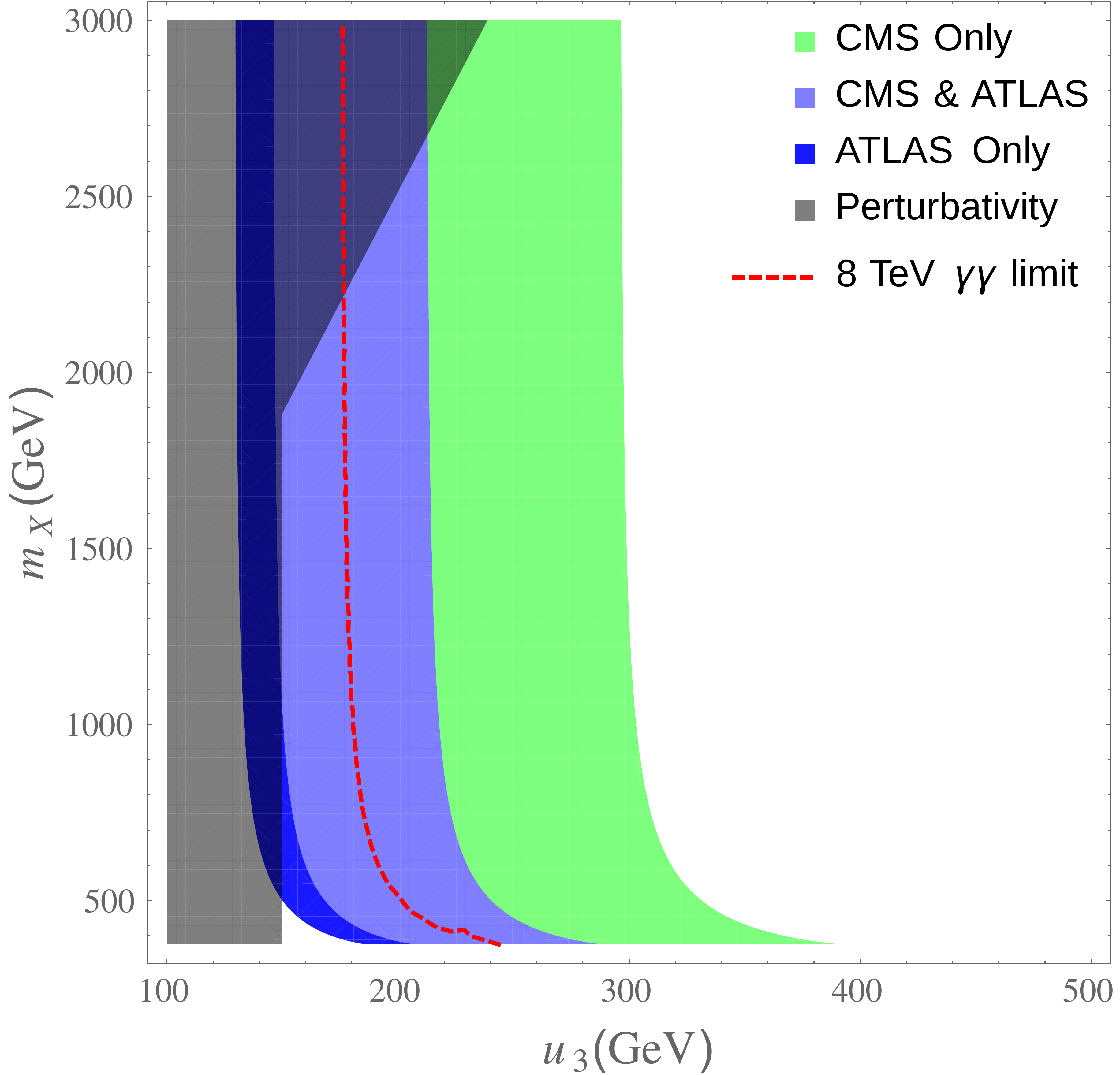}
\caption{\label{fig6} The allowed $m_X$ - $u_3$ range (for $\kappa_{\chi_2} = 0.5$) corresponding to CMS (green), ATLAS (deep blue) and the overlap (light blue) ranges  with $95\%$ confidence level. Also, shown is the $95\%$ confidence level $H_1 \to \gamma \gamma$ exclusion line (red dashed) from 8 TeV run with the regions on the left of the line being incompatible with it. The black shaded region is also excluded by the perturbativity constraints.} 
\end{figure}

     
  In this case, in addition to the decay channels discussed in previous section, $H_1$ decay to dark matter is also kinematically allowed and it can have appreciable invisible decay width. In Fig.~\ref{fig6} we show the allowed parameter range for the exotic quark masses and $u_3$, that can explain the observed 750 GeV diphoton excess for both the CMS and ATLAS experiments within $95 \%$ confidence level. In plotting Fig.~\ref{fig6} we have taken $\kappa_{\chi_2} = 0.5$ and have also imposed all the constraints from 8~TeV run listed in Eq.~\ref{8tevcon}.     \\ 
     
As in the previous case, again the only significant constraint from 8~TeV run comes from the $\gamma \gamma$ decay channel which is also plotted in Fig.~\ref{fig6}. The parameter space on the left of the red line is incompatible with the 8 TeV data. As before, the constraints from other channels including the invisible decay to dark matter are rather weak and do not give any additional constraint.  As an example the values for these decay channels for a benchmark point ($m_X = 1$ TeV and $u_3 = 180$ GeV) on the $\gamma \gamma $ exclusion line of Fig.~\ref{fig6} are given as  

\begin{eqnarray}
   \sigma \times \rm{Br}(H_1 \rightarrow \gamma \gamma) & = & 1.5 \, \rm{fb}, \qquad 
     \sigma \times \rm{Br}(H_1 \rightarrow g g) \, = \, 409 \, \rm{fb},  \nonumber \\
     \sigma \times \rm{Br}(H_1 \rightarrow Z \gamma) & = & 0.89 \, \rm{fb}, \qquad 
     \sigma \times \rm{Br}(H_1 \rightarrow Z Z) \, = \, 0.14 \, \rm{fb}. \nonumber \\
\sigma \times \rm{Br}(H_1 \rightarrow \chi_2 \chi^*_2) & = & 244 \, \rm{fb}
  \label{dmexc}
 \end{eqnarray}

As clear from Eq.~\ref{dmexc}, like the previous case here also only the constraints from $\gamma \gamma$ channel for 8~TeV run are important. The constraints from all other channels are comfortably satisfied. Furthermore, just like the previous case, in this case also our model has enough parameter space compatible with both the observed 13~TeV excess and the 8~TeV constraints. Thus the 750 GeV excess can be understood in our model as the decay of $H_1$ to a pair of
photons. \\

Finally, before ending the section we like to discuss briefly about the total decay width of $H_1$ for the two cases. The first thing to note is that given the current low statistics, the estimates of decay width are very poor. This aspect is highlighted by the fact that while CMS data prefers narrow decay width of around a few GeV for the resonance, the ATLAS prefers a relatively broader resonance with decay width $\sim 45$ GeV.  Thus the current estimates of decay width are highly uncertain and are likely to change significantly in the future runs. \\

In our model, if the $H_1$ decay to dark matter is kinematically forbidden then the dominant decay channels will all be loop induced with $H_1 \to g g$ being the most significant. In such a scenario $H_1$ will be a narrow resonance with total decay width up to a few GeVs. However, if $H_1$ decay to dark matter is kinematically allowed then it can have significant invisible decay width owing to the fact that such a decay is not loop suppressed. In this case the $H_1$ can be a broad resonance. \\

If in future runs the ATLAS experiments estimates of a broad resonance persists then for our model it will imply a significant invisible decay width. Depending on the value of $\kappa_{\chi_2}$, $H_1$ can have decay width upto 40 GeV, albeit for a small parameter range. In such a case, a better solution can be obtained by adding a pair of $SU(2)_L$ singlet charged leptons to our model. However, at this stage we feel that such an extension of our model is premature and not necessary.


\section{Conclusions}
\label{sec5}  


The recently observed $750$ GeV diphoton excess has drawn significant attention  as it could  be the first signs of new physics at LHC. Although it is too early to conclude 
that this is a definite sign of new physics but nonetheless it raises an intriguing possibility that it might originate from decay of a hitherto unknown particle.  In this work we have looked at the extended gauged $B - L$ symmetry model with unconventional charges for the right handed neutrinos as a possible candidate new physics model to explain the $750$ GeV excess. The model was originally constructed to obtain Dirac neutrinos with naturally small masses and also has a long lived dark  matter particle. Unlike the conventional gauged $B - L$ symmetry model where the $B - L$ scale is expected to be quite high, being related with the seesaw scale, in our model the $B - L$ scale can be well within the LHC range thus opening up the possibility of testing its various aspects at LHC.  \\

We have looked at the possibility that the observed diphoton excess can arise due to decay of the scalar particle $H_1$ into two photons. This scalar in our model is predominantly composed of the singlet scalars $\chi_3$ and $\chi_6$ which are essential ingredients of the model. They are required in order to spontaneously break the gauged $B - L$ symmetry as well as to obtain Dirac neutrinos with small masses. We have further shown that the model not only explains the diphoton excess but also satisfies all the other experimental constraints like non-observation of any excess in dilepton, dijet, diboson and invisible channels. It also has a $125$ GeV particle $h$ which has almost SM Higgs like couplings to the other SM particles and satisfies all the other experimental constraints for the $125$ GeV scalar. Moreover, since in our model $h$ is predominantly composed of the $SU(2)_L$ doublet scalar with very small admixture from $SU(2)_L$ singlet scalars, it naturally explains why the $125$ GeV particle has almost SM like 
couplings. \\

Thus to conclude, the gauged $B - L$ model considered here appears to be a promising candidate for new physics. It has all the right ingredients to explain not only the $750$ GeV diphoton excess but all the other experimental results both for the $750$ GeV resonance as well as the $125$ GeV resonance. Moreover, the model also connects the observed new physics 
with the already well know and long standing problems of neutrino masses and dark matter and attempts to provide a unifying solution to all of them. Also, it has several testable predictions like existence of heavier particles in $\sim 1$ TeV range, Dirac nature of neutrinos and candidate for dark matter. These aspects can be tested in future run of LHC as well as in dark matter direct detection experiments and in various neutrino physics experiments.  


\begin{acknowledgments}

We will like to thank Ernest Ma for his valuable comments and suggestions which were of immense help in successful completion of this work.
 
\end{acknowledgments}


\end{document}